\documentclass{article}
\date{}
\begin{document}
\title{Relativistic diffusing particles, entropy and varying cosmological  constants}
\author{Z. Haba\\Institute of Theoretical Physics, University of Wroclaw, 50-204 Wroclaw\\
Plac Maxa Borna 9, Poland }

\date{}
\maketitle
\begin{abstract}
We discuss  Einstein equations with a non-conserved energy-momentum
$T^{\mu\nu}$.

The energy-momentum tensor $\tilde{T}^{\mu\nu}$ on the right hand side
of Einstein equations is expressed in the form $8\pi
GT^{\mu\nu}-g^{\mu\nu}\Lambda+\Sigma^{\mu\nu}$ where $G$ is the
Newton constant, $\Lambda$ is the cosmological term and $T^{\mu\nu}$
is the energy-momentum of an ideal fluid.
 The remaining part $\Sigma^{\mu\nu}$ describes
the viscosity and a heat flow.
  We discuss  space-time-dependent
Newton ``constant" $G$ and  the cosmological term $\Lambda$.  The
covariant conservation law for $\tilde{T}^{\mu\nu}$ imposes a
relation between the energy-momentum (non)conservation and the
variation of $G$ and $\Lambda$.  The covariant derivative
$\nabla_{\mu}T^{\mu\nu}$ can be related to the first law of
thermodynamics  and expressed by the temperature $T$ and the
entropy $S$.   The energy-momentum non-conservation in our models is a consequence of a diffusion
of its constituting particles.
 We
describe a gas  of diffusing massive particles in a homogeneous
isotropic space-time as a system with a time-dependent linear
barotropic equation of state
 $P_{t}=w_{t}\rho_{t}$ with $0<w_{t}\leq \frac{1}{3}$.
 The gas of massive diffusing particles in an expanding universe tends in a limit
of infinite time to the gas of diffusing massless particles.
   We obtain results on the energy density, entropy and phase space dstribution
   in a soluble model of massless relativistic particles in a spatially flat expanding universe. 
\end{abstract}
 \maketitle
\section{Introduction}
The geometry of the universe in Einstein general relativity is
determined by the energy-momentum tensor $\tilde{T}^{\mu\nu}$
($\mu,\nu=0,1,2,3$, we use the conventions of \cite{AMT})
\begin{equation} G^{\mu\nu}=\tilde{T}^{\mu\nu},
\end{equation}where $G^{\mu\nu}=R^{\mu\nu}-\frac{1}{2}g^{\mu\nu}R$ ($R^{\mu\nu}$
is the Ricci tensor).
 The standard Einstein equations can be expressed in
the form (velocity of light $c=1$)
\begin{equation}
G^{\mu\nu}=\tilde{T}^{\mu\nu}\equiv 8\pi G T^{\mu\nu}-g^{\mu\nu}\Lambda+\Sigma^{\mu\nu}.
\end{equation}
 $G$ is the Newton constant. The tensor
$T^{\mu\nu}$ on the rhs of eq.(2) is usually assumed in the form
of an ideal fluid. $\Lambda$ is the cosmological term.
$\Sigma^{\mu\nu}$ describes viscosity and a heat flow in the fluid
\cite{israel}\cite{wert}. Such a term is usually discarded (see
however \cite{viscosity}). The system (2) with constant $G$ and
$\Lambda$ (without $\Sigma^{\mu\nu}$) constitutes  the basis of the
standard $\Lambda$CDM model. There are some attempts to extend the
$\Lambda$CDM model ( see \cite{desi}) to varying cosmological
constants in order to explain some problems in this model (the
need for dark matter, coincidence problem, $H_{0}$-tension
\cite{skara}). The coincidence problem could be approached by an
assumption that the baryonic matter, dark matter and dark energy
are related by an interaction. In refs.\cite{hss}\cite{habagrg} we
discussed a model with a diffusing dark matter.  In order to work with a conserved
energy-momentum $\tilde{T}^{\mu\nu}$ (with constant $G$ and
$\Lambda$)  we had to add a compensating energy-momentum for the
dark energy.

 In this paper we discuss Einstein equations for a system of particles
 (baryons and dark matter) moving  in a passive medium.
  The non-conserved energy-momentum can appear in Einstein equations if the Newton ``constant"
and (or) the cosmological term are space-time dependent. We
consider Einstein general relativity of a   fluid of
relativistic particles described by a phase space distribution
$\Omega$. The phase space distribution defines the energy-momentum
entering the Einstein equation and the particle current applied
for a decomposition of the energy-momentum in its ideal fluid and
a dissipative part.
However, in prospective studies we  would
apply the phase space distribution in cosmology for a description
of formation and
 dynamics of clusters of galaxies. The energy-momentum is not
conserved if some degrees of freedom in $\Omega$ are ignored ( so
that there is just an averaging over these degrees of freedom).
We obtain a dissipative system.
The Markov approximation leads to a diffusive description of the neglected degrees of freedom.
Such a description is well-known from the theory of Brownian motion. The Brownian particles are
described in statistical physics by their probability distribution. A detailed knowledge of the neglected
degrees of freedom (the medium) is not needed. The medium just contributes to the diffusion constant.
The (incomplete)  energy-momentum $T^{\mu\nu}$ on the rhs of Einstein equations (1)-(2) 
does not contradict Bianchi identities if the Newton ``constant"
$G$ and (or) the cosmological term $\Lambda$ are space-time
dependent. Space-time dependence of $G$ and $\Lambda$ in Einstein equations 
has been discussed earlier in \cite{dicke}\cite{bri2}\cite{ong}\cite{ong2}\cite{uzan}\cite{gupta}\cite{blanchard}\cite{cooperstock}
\cite{v1}\cite{v2}\cite{v3}\cite{v4}. In theoretical models the space-time dependence of cosmological
constants can result from an addition of some other fields to the model (as in \cite{dicke}) or
from quantum effects incorporated into the classical model ( as vacuum expectation value of the energy-momentum tensor or renormalization of the Newton constant).
The variation of the
 Newton ``constant" $G$  can also be explained in the phase space
 picture as a presence of some unobservable particles which
 interact with the baryonic and dark matter.
  The fluid $\tilde{T}^{\mu\nu}$ may
consist of conserved matter (which does not interact with the rest
of the system) and the dark matter diffusing in an unspecified medium. The particles of this medium do not enter the energy-momentum
$\tilde{T}^{\mu\nu}$ but their presence is experienced as a
diffusion of the dark matter and a variation of cosmological
constants.

Diffusing matter has been introduced in Einstein gravity in some
earlier papers .
 A modification of Einstein gravity by an  energy-momentum tensor of a
diffusion (resulting from a discreteness of space-time) has been
discussed in \cite{perez2}\cite{perez1} \cite{perez5}\cite{cruz} \cite{perez4}. 
This model can be considered as a version of unimodular gravity \cite{perez4} which 
allows energy-momentum non-conservation if the covariant derivative of the energy-momentum is
related to the cosmological term. In our approach
in order to preserve the
general covariance we consider the relativistic diffusion. The
relativistic diffusion is a well-studied model
\cite{dudley}\cite{franchi}\cite{habapre}\cite{alcantara}. Its
application to cosmology has been considered in
\cite{hss}\cite{habacqg}\cite{benisty}\cite{calo1}\cite{calo2}
\cite{london1}\cite{london2}\cite{maity}\cite{sahlu}\cite{recent}. In
\cite{perez2}\cite{perez1}\cite{cruz} the model has been applied to an
explanation of dark energy and the $H_{0}$ tension.
  In such a model
the dark energy results from a non-conservation of the energy of
the diffusive  matter. This is a certain modification of
the model 
\cite{hss}\cite{habagrg}\cite{habacqg} \cite{maity}\cite{sahlu}\cite{recent}\cite{sahlu2} where the dark
energy has its own energy-momentum in order to compensate for the
dark energy-momentum non-conservation. The phenomena with a
violation of the energy-momentum conservation cannot be described
by Einstein equations with fixed cosmological constants . We must
either compensate $\nabla_{\mu}T^{\mu\nu}$ by another
(non-conserved) energy-momentum of the dark energy or admit
varying cosmological constants ( a varying $\Lambda$ in Einstein gravity
with a diffusion has been considered in \cite{calo1}).

 In this paper we
study varying constants and a non-conservation of the
energy-momentum $T^{\mu\nu}$ of an ideal fluid. In sec.2 we obtain
the Bianchi identities resulting from a conservation of
$\tilde{T}^{\mu\nu}$(1) and non-conservation of $T^{\mu\nu}$ (2).
A covariant conservation of the total
energy-momentum  $\tilde{T}^{\mu\nu}$ (2) (including the varying
$G$ and $\Lambda$) leads to equations relating the Newton
``constant" and (or) the cosmological term $\Lambda$ 
to $\nabla_{\mu}T^{\mu\nu}$. 
In sec.3 we introduce a model of a gas of relativistic diffusing
particles in an expanding universe. The thermodynamics of such a
gas is discussed in sec.4. In sec.5 we consider a J\"uttner-type
solution for diffusing relativistic massless particles. The model
of massless particles can be considered as a large time limit of
the massive one. In this model we study the effect of diffusing
particles on the possible variation of $G$ and $\Lambda$.According
to the first law of thermodynamics $\nabla_{\mu}T^{\mu\nu}$ is
related to the heat flow. Then owing to the second law of
thermodynamics the heat flow should be expressed by the change of
entropy $TdS$ (where $T$ is the temperature). In sec.5 applying
the thermodynamics we express the variation of cosmological
constants by the entropy $S$ of a gas of diffusing particles ( in
particular, we obtain $\ln G \simeq S$). In sec.6 we describe the
entropy of a diffusing gas in more detail. We show that the
microscopic definition of the entropy coincides with the
thermodynamic definition. In sec.7 we discuss solutions of
the
diffusion equation in a fixed metric and subsequently the Einstein
equations which determine this metric. In the Appendix  we
study in detail the general solution of the massless diffusion
equation in an expanding metric.
In particular, we discuss  the
exponential expansion of the universe and the coasting cosmology
\cite{kolb}  which appear as distinguished solutions of Einstein
and diffusion equations. A solution of the diffusion equation 
together with Einstein equations 
describes a distribution of particles (galaxies) in the phase
space defined by the universe evolution. 
\section{Einstein equations and Bianchi identities }

 The Bianchi identity applied to eq.(1) leads to the equation
\begin{equation}
\nabla_{\mu}\tilde{T}^{\mu}_{\nu}=0.
\end{equation} If  we can define a time-like
fluid four-velocity $u^{\mu}$ (satisfying the normalization
$u_{\mu}u^{\mu}=-1$) then we can decompose any energy-momentum
tensor in the form \cite{israel}\cite{wert}
\begin{equation}\begin{array}{l}
\tilde{T}^{\mu\nu}=8\pi G
T^{\mu\nu}+\Sigma^{\mu\nu}-g^{\mu\nu}\Lambda\equiv \hat{T}^{\mu\nu}-g^{\mu\nu}\Lambda
\end{array}\end{equation}
where (ideal fluid)
\begin{equation}
T^{\mu\nu}=(\rho+P)u^{\mu}u^{\nu}+g^{\mu\nu}P.
\end{equation}
Then,  $\rho$ has the meaning of the energy-density and $ P$  is
the pressure.

We can express $\rho$ , $P$ and $\Sigma^{\mu\nu}$ by suitable projections of the energy-momentum tensor
$\tilde{T}^{\mu\nu}$ of eq.(4). These projections give
\begin{equation}
8\pi G\rho+\Lambda =u^{\mu}\tilde{T}_{\mu\nu}u^{\nu},
\end{equation}
\begin{equation}
8\pi GP-\Lambda= \frac{1}{3}\tilde{T}_{\mu\nu}\triangle^{\mu\nu}
\end{equation}
with
\begin{equation}
\triangle_{\alpha}^{\mu}=\delta^{\mu}_{\alpha}+u^{\mu}u_{\alpha}.
\end{equation}
Then,
\begin{equation}
\Sigma^{\mu\nu}=I^{\mu}u^{\nu}+I^{\nu}u^{\mu}+\tilde{\Pi}^{\mu\nu}
\end{equation}
$I^{\mu}$ has the meaning of the energy flow. It is subtracted from $\Sigma^{\mu\nu}$
as
\begin{equation}
I^{\mu}=u_{\nu}\tilde{T}^{\alpha\nu}\triangle_{\alpha}^{\mu}.
\end{equation}
where the tension tensor
$\tilde{\Pi}^{\mu\nu}$ is separated from $\Sigma^{\mu\nu}$ by the condition
\begin{equation}
\tilde{\Pi}^{\mu\nu}u_{\mu}= \tilde{\Pi}^{\mu\nu}g_{\mu\nu}=0.
\end{equation}

The Bianchi identity (3) gives an equation for the ``fluids"
\begin{equation}
8\pi
\partial_{\mu}G T^{\mu}_{\nu}+8\pi
G\nabla_{\mu}T^{\mu
}_{\nu}-\partial_{\nu}\Lambda+\nabla_{\mu}\Sigma^{\mu }_{\nu}=0.
\end{equation}
We decompose the Bianchi identities (12) into the parts parallel
to the velocity (through a multiplication of eq.(12) by $u^{\mu}$ ) and orthogonal to the velocity
(through a multiplication by $\triangle_{\mu\alpha}$). The parallel part
is
\begin{equation}\begin{array}{l}
-8\pi u^{\mu}\partial_{\mu}G\rho-8\pi Gu^{\mu}\partial_{\mu}\rho
-u^{\nu}\partial_{\nu}\Lambda-8\pi
G(\rho+P)\nabla_{\mu}u^{\mu}+u_{\nu}\nabla_{\mu} \Sigma^{\mu\nu}=0
\end{array}\end{equation}
The orthogonal part reads
\begin{equation}\begin{array}{l}
8\pi P\partial^{\mu}G\triangle_{\mu\alpha}+8\pi
G\partial^{\mu}P\triangle_{\mu\alpha}-\triangle_{\nu\alpha}\partial^{\nu}\Lambda+8\pi
G(\rho+P)u^{\mu}\nabla_{\mu}u^{\nu}\triangle_{\nu\alpha}\cr+\triangle_{\nu\alpha}\nabla_{\mu}\Sigma^{\mu\nu}=0.
\end{array}\end{equation}
In general, it is difficult to solve the Bianchi identities. Note
however, that if
$\partial_{j}P=\partial_{j}G=\partial_{j}\Lambda=0$ then in the
comoving frame the orthogonal  Bianchi identities are satisfied.
In the next section we discuss relativistic systems with a
phase-space distribution $\Omega(x,p)$. If the dependence of
$\Omega$ on the momentum $p$ is in the form $p_{\mu}u^{\mu}$ with
a cetrtain four-vector $u$ then the energy-momentum is in the form
of an ideal fluid (5) (for a general phase space distribution it
will have the form(4)).

We assume that the Riemannian manifold is globally hyperbolic.
Then, we can introduce coordinates \cite{global} such that the
metric takes the form
\begin{equation}
dl^{2}=g_{\mu\nu}dx^{\mu}dx^{\nu}= g_{00}
dt^{2}+\gamma_{jk}dx^{j}dx^{k}=-\vert g_{00}\vert
dt^{2}+\gamma_{jk}dx^{j}dx^{k}.
\end{equation}

We shall discuss models with a particle current (a stream of
particles) $N^{\mu}$. Such a current exists if the particles have
a phase space distribution ( sec.3). Then, Eckart's \cite{eckart}
definition of the fluid velocity is
\begin{equation} u^{\mu}=(-N_{\nu}N^{\nu})^{-\frac{1}{2}}N^{\mu}.
\end{equation}
A solution of Einstein equations describing  an expanding
isotropic homogeneous universe (with the curvature index
$k=-1,0,1$) has the  FLRW metric
\begin{equation}
dl^{2}=-dt^{2}+a^{2}\Big((1-kr^{2})^{-1}dr^{2}+r^{2}(d\theta^{2}+\sin^{2}\theta
d\phi^{2})\Big).
\end{equation}
In the metric (17) the fluid velocity satisfies
$\nabla_{\mu}u^{\mu}=(-g)^{-\frac{1}{2}}\partial_{\mu}((-g)^{\frac{1}{2}}u^{\mu})=3H$
(where $g= \det( g_{\mu\nu})$ and $H=a^{-1}\partial_{t}a$). For
the ideal fluid we have
\begin{equation}
u^{\nu}\nabla_{\mu}T^{\mu}_{\nu}=-(\rho+P)\nabla_{\mu}u^{\mu}+u^{\mu}\nabla_{\mu}\rho.
\end{equation}
 Hence, from eqs.(13) and (18) under the
assumption that $\rho$ and $p$ depend only on time
\begin{equation} 8\pi
\partial_{t}(G\rho) +24\pi GH(\rho+P)+\partial_{t}\Lambda=0
\end{equation}If $P=w\rho$ then eq.(19) is an equation for
$G\rho$. However, if the relation between $\rho$ and $P$ is
non-linear (e.g.,  a polytropic gas) then $G$ and $\rho$ enter
separately Einstein equations and Bianchi identities.

 $T^{\mu\nu}$ may be  a sum of many fluids with some conserved
parts in it
\begin{equation} \rho=\rho^{c} +\tilde{\rho},
\end{equation}
where by the index ``c" we  mean a fluid with a conserved
energy-momentum
\begin{equation}\partial_{t}\rho^{c}
+3H(\rho^{c}+P^{c})=0.\end{equation} If
$\tilde{P}=\tilde{w}\tilde{\rho}$ and $P^{c}=w^{c}\rho^{c}$ then
eq.(19)   depends separately on $\rho$ and $G$
\begin{equation} 8\pi
\partial_{t}(G\rho)+24\pi H(1+w^{c})G\rho^{c}+24\pi H(1+\tilde{w})G(\rho-\rho^{c})+\partial_{t}\Lambda=0
\end{equation}

Eq.(14) is satisfied if  $P$ depends only on time as  (in the
comoving frame $u=(1,{\bf 0})$) , $T^{0j}=0$ and
$\nabla_{j}T^{j}_{k}=\partial_{k}P=0$.

The Einstein
 equations (1) in the metric (17)can be decomposed in the comoving frame   in the $00$ and $\mu j$
parts. The $G^{00}$ part is
\begin{equation}
    (a^{-1}\frac{da}{dt})^{2}=- \frac{k}{a^{2}}+G\frac{8\pi}{3} \rho
    +\frac{1}{3}\Lambda.
\end{equation}
Adding the $G^{00}$ equation to the $G^{kj}$ equation we obtain
the second (Friedmann) equation
\begin{equation}
a^{-1}\frac{d^{2}a}{dt^{2}}=-G\frac{4\pi}{3}
(\rho+3P)+\frac{1}{3}\Lambda.
\end{equation}
The second Friedmann equation (24) follows from the first equation
(23) and the conservation law (19)(even if $G$ and $\Lambda$ are
time dependent).

\section{Non-conserved energy-momentum:relativistic diffusion}
In this section we consider a gas of particles moving on a
(pseudo)Riemannian manifold and described by a phase space
distribution $\Omega$. We concentrate on a model of relativistic 
particles diffusing in a medium expanding in the
metric (15). Let us  briefly review the model  which may have a wider application in
astrophysics and cosmology
\cite{hss}\cite{habagrg}\cite{perez2}\cite{perez1}\cite{perez5}\cite{habapre}\cite{habacqg}\cite{calo1}\cite{calo2}\cite{london1}\cite{london2}.
A gas of particles is described by a  distribution $\Omega $ on
the phase space \cite{stewart}\cite{vega}\cite{andrea}\cite{ehlers}. If
the particle's dynamics (geodesics) are determined by classical
evolution equations in a gravitational field, then the
energy-momentum conservation law is a consequence of the Liouville
equation \cite{stewart}(where $\Gamma^{\mu}_{\nu\rho}$ are the
Christoffel symbols)
\begin{equation}
\begin{array}{l}
(p^{\mu}\partial^{x}_{\mu}-\Gamma^{k}_{\mu\nu}p^{\mu}p^{\nu}\partial^{p}_{k})\Omega=
0,
\end{array}\end{equation}
where $\partial^{p}$ denotes derivatives over the momenta and
$\partial^{x}$ is a derivative over a space-time coordinate $x$.
The Greek indices run from $0$ to $3$, Latin indices denoting
spatial components have the range from $1$ to $3$. The
energy-momentum tensor is defined by
\begin{equation}
\hat{T}^{\mu\nu}=\sqrt{g}\int d{\bf p}p_{0}^{-1}p^{\mu}p^{\nu}\Omega,
\end{equation}
where  $p_{0}$ is determined from the mass-shell condition
$p_{\mu}p^{\mu}=-m^{2}$ ($m$ is the particle's mass).We can also define the conserved particle current
($\nabla_{\mu}^{x}N^{\mu}=0$)

\begin{equation}
	N^{\mu}=\sqrt{g}\int d{\bf p}p_{0}^{-1}p^{\mu}\Omega.
\end{equation}
 $\hat{T}^{\mu\nu}$ in general
is not of the form of an ideal fluid (5) but has a decomposition (4)-(9). However, 
in subsequent sections we shall consider solutions of the diffusion equation which lead to the
energy-momentum tensor of the form of the ideal fluid (5).

We could consider a more general phase space distribution
\begin{displaymath}
\hat{\Omega}_{t}({\bf x},{\bf p},{\bf l})=\Omega_{t}({\bf x},{\bf
p})\omega_{t}({\bf x},{\bf l })
    \end{displaymath}
describing different sets of particles with momenta ${\bf p}$ and
${\bf l}$. If we do not include the $l^{\mu}l^{\nu}$ integrals of
eq.(26) into the energy-momentum tensor $\tilde{T}^{\mu\nu}$ for
the phase space distribution $\hat{\Omega}_{t}$  then their sole
effect is the multiplicative factor $G\rightarrow G\int d{\bf
l}\omega_{t}({\bf x},{\bf l})$ in the energy-momentum tensor (26). In
general, without the inclusion of the ${\bf l}$- particles the energy-momentum tensor (26)
will be not conserved.

 We preserve the formulas (26)-(27) for a diffusion
which still is described by a phase space distribution $\Omega_{t}
({\bf x},{\bf p})$. The diffusion can be considered as a particle motion along random geodesics .
 The diffusion takes place on the
hyperboloid  ${\cal H}_{+}$ (the positive energy mass-shell ; see
\cite{dudley}\cite{franchi}\cite{habapre}\cite{habacqg}\cite{calo2}). $N^{\mu}$ (27) is a time-like vector which can be
related to a fluid velocity current $N^{\mu}=Nu^{\mu}$ according
to eq.(16) .
The transport equation is \begin{equation}
\begin{array}{l}
(p^{\mu}\partial^{x}_{\mu}-\Gamma^{k}_{\mu\nu}p^{\mu}p^{\nu}\partial^{p}_{k})\Omega=
\kappa^{2}\triangle_{H}\Omega,
\end{array}\end{equation} where $\kappa^{2}$ is the diffusion
constant.

 The operator $\triangle_{H}$ is the
Laplace-Beltrami operator on the mass hyperboloid ${\cal H}_{+}$
\begin{displaymath}
g_{00}p^{0}p^{0}+\gamma_{jk}p^{j}p^{k}=-m^{2}.
\end{displaymath}
The metric is defined by
\begin{equation}
ds^{2}=m^{-2}g_{\mu\nu}dp^{\mu}dp^{\nu}\equiv h_{jk}dp^{j}dp^{k},
\end{equation} where $p^{0}$ is expressed by spatial components
 \begin{equation}p^{0}=\sqrt{\vert
g_{00}\vert^{-1}(m^{2}+\gamma_{jk}p^{j}p^{k})}.
\end{equation}
The metric is
\begin{equation}
h_{jk}=m^{-2}(\gamma_{jk}-p_{j}p_{k}\omega_{g}^{-2}),
\end{equation}
where $p_{j}=\gamma_{jk}p^{k}$ and

\begin{equation}
\omega_{g}^{2}=m^{2}+\gamma_{jk}p^{j}p^{k}.
\end{equation}
 Then
\begin{equation}
\triangle_{H}=\frac{1}{\sqrt{h}}\partial^{p}_{j}h^{jk}\sqrt{h}\partial^{p}_{k},
\end{equation}
where
\begin{equation}
h^{jk}=m^{2}\gamma^{jk}+p^{ j}p^{ k},
\end{equation}  $\partial^{p}_{j}=\frac{\partial}{\partial
p^{ j}}$ and
$h=\det(h_{jk})=m^{-4}\det(\gamma_{jk})\omega_{g}^{-2}$ is the
determinant of the metric $h_{jk}$.

 If $\kappa\neq 0$ then  the energy-momentum tensor is not conserved but
\cite{calo2} \cite{habampa}
\begin{equation}
\nabla_{\mu}^{x}\hat{T}^{\mu\nu}=3\kappa^{2}N^{\nu}.
\end{equation}

Now eq.(12) reads (where $T^{\mu\nu}\rightarrow\hat{T}^{\mu\nu}$)
\begin{equation}
8\pi
\partial_{\mu}G \hat{T}^{\mu}_{\nu}+24\pi
G\kappa^{2}N_{\nu}-\partial_{\nu}\Lambda=0.
\end{equation}
In our model the particle current is conserved (as a consequence
of eq.(28))\begin{equation}
\nabla_{\mu}N^{\mu}=(-g)^{-\frac{1}{2}}\partial_{\mu}((-g)^{\frac{1}{2}}N^{\mu})=0.
\end{equation}
This means that the particle number is constant
\begin{equation}\begin{array}{l}
\partial_{t}\int d{\bf
x}((-g)^{\frac{1}{2}}N^{0})=-\int d{\bf
x}\partial_{j}(-g)^{\frac{1}{2}}N^{j})=0\end{array}
\end{equation}
 In the comoving frame only the temporal
component of the current is not zero
\begin{equation}N^{0}=n(-g)^{-\frac{1}{2}},\end{equation}where
$n$ may be space dependent.

 The
zeroth component of eq.(36) reads ( assuming $G$
and $\Lambda$ depend only on time)
\begin{equation}
8\pi
\partial_{t}G \hat{T}^{0}_{0}+24\pi
G\kappa^{2}ng_{00}(-g)^{-\frac{1}{2}}-\partial_{t}\Lambda=0.
\end{equation}

 From now on
we restrict ourselves to a flat expanding metric ($k=0$ in FLRW
metric (17))
\begin{equation}
dl^{2}=-dt^{2}+ a^{2}d{\bf x}^{2}
\end{equation}

Writing $ \nabla_{\mu}\hat{T}^{\mu}_{0}=3\kappa^{2}N_{0}$  in terms of
$\rho$ and $P$ we obtain

\begin{equation}
\partial_{t}\rho
+3H(\rho+P)=3\kappa^{2}n(-g)^{-\frac{1}{2}}.
\end{equation}
This equation can be solved for $\rho$ if $P=w\rho$. Then, we can
express the solution of
 eq.(42)as
 \begin{equation}\begin{array}{l}
 \rho(t)=a^{-3(1+w)}\Big(\rho(t_{0})a(t_{0})^{3(1+w)}
 +3n\kappa^{2}\int_{t_{0}}^{t}a(s)^{3w}ds\Big)\cr=3n
 a^{-3(1+w)}(T_{0}+\kappa^{2}\int_{t_{0}}^{t}a(s)^{3w}ds),
 \end{array}\end{equation}where
we have defined the initial conditions by means of a parameter
$T_{0}$ ( which will be related to the initial or present time
temperature at $t_{0}$, see sec.4) in such a way that
\begin{displaymath}
\rho(t_{0})=3na_{0}^{-3(1+w)}T_{0}.
\end{displaymath}

In a way similar to the standard (non-relativistic) Brownian motion with a friction we can ascribe a temperature $T$
to the gas of diffusing particles \cite{mackey}.
 The formula for the temperature follows from eq.(43)
 (\cite{habagrg} and sec.4, eq.(60))

 \begin{equation}
 T=a^{-3w}(T_{0}+\kappa^{2}A).
 \end{equation}where
\begin{displaymath}
A=\int_{t_{0}}^{t}a(s)^{3w}ds.\end{displaymath} Note that
\begin{displaymath}
\frac{dT}{dt}=-3wHT+\kappa^{2}.
\end{displaymath}
For an exponential expansion $a=\exp(Ht)$ we have
\begin{displaymath}
T=\exp(-3wHt)\Big(T_{0}-\frac{\kappa^{2}}{3wH}\exp(3wHt_{0})\Big)+\frac{\kappa^{2}}{3wH}
\end{displaymath}
For a power-law expansion $a=t^{\alpha}$ we get
\begin{displaymath}\begin{array}{l}
T=(T_{0}-\frac{\kappa^{2}}{1+3w\alpha}t_{0}^{1+3w\alpha})t^{-3w\alpha}+\frac{\kappa^{2}}{1+3w\alpha}t.
\end{array}\end{displaymath}
Hence, for an exponential expansion the temperature of the fluid
is decaying (from its  value $\exp(-3Hwt_{0})T_{0}$) tending to a constant with the speed $a^{-3w} $ when $t\rightarrow
\infty$ . For a power-law expansion it is
decaying from $T_{0}t_{0}^{-3w\alpha}$ at small $t\geq t_{0} $  with the rate $a^{-3w}$ (in agreement
with the adiabatic thermodynamics of sec.4) and subsequently
linearly growing to infinity for a large time.

 Let us consider
now the universe evolution determined by eq.(41) and the Bianchi
identity resulting from eqs.(12) and (40)
\begin{equation}
a^{-3(1+w)}(T_{0}+\kappa^{2}A)\partial_{t}G+G\kappa^{2}a^{-3}=-\frac{1}{24n\pi}\partial_{t}\Lambda
\end{equation}
If $G=const $ then the Bianchi identity (45), the expression of
the energy-momentum non-conservation  (42) and the first Friedmann
equation(23) constitute a closed system of first order
differential equations for $\Lambda,\rho, a$ with the unique
solution under given initial conditions ( in particular, we may
assume for $\Lambda$ zero or non-zero initial condition). If
G=const then it follows (this is an equation of Perez et al
\cite{perez2} and Calogero \cite{calo1})
\begin{equation}
\Lambda=\Lambda_{ex}-24\pi nG\kappa^{2}\int_{t_{0}}^{t}a^{-3}
\end{equation}
When we insert $\rho$ (43) and $\Lambda$ (46) at $G=const$ in the
first Friedman equation (23) then we obtain an integro-differential
equation which is difficult to solve exactly. At small $\kappa^{2}$ we have the standard (non-diffusive)
behaviour of the solutions of the Friedmann equations. $\Lambda_{ex} $ in our model is just an arbitrary parameter in the initial condition.
If $\Lambda_{ex}>0$
then the  Friedmann equations will have solutions with the asymptotic behaviour $a\simeq
\exp(t\sqrt{\frac{\Lambda_{ex}}{3}})$.  If $\Lambda_{ex}=0$ then
asymptotically $a\simeq t$ at large time (coasting cosmology \cite{kolb}). This is different from
$t^{\frac{2}{3(1+w)}}$ for the evolution without diffusion. If we admit $G$ to depend on time then
in order to have the unique solution we must fix one of the
functions ( possibly from astronomical observations) $a$,$\rho$,
$\Lambda$ or $G$ (this still will be discussed in sec.7).

If $\Lambda=const$ and $\rho^{c}=0$ (under the assumption
$P=w\rho$) then the solution of eq.(22) is $G\rho=Ka^{-3(1+w)}$
 . Hence, the diffusion has no effect on
the first Friedmann equation ( we obtain the standard solution for
the Friedmann equation). However, according to eq.(45)
(owing to the additional equation (35))\begin{equation} G=G_{0}T_{0}(T_{0}+\kappa^{2}A)^{-1}.
\end{equation}
If there is to be a modification of the gravitational potential
resulting from the source $G\rho$ then we must assume a varying
$\Lambda$. In general,
from eq.(45)
\begin{equation}\begin{array}{l}
G=G_{0}T_{0}(T_{0}+\kappa^{2}A)^{-1}-\frac{1}{24\pi n}
(T_{0}+\kappa^{2}A)^{-1}\int_{t_{0}}^{t}a^{3(1+w)}\partial_{s}\Lambda
ds.
\end{array}\end{equation}
The solution $\Omega$ of the diffusion equation
defines $\rho$ and $P$ as a function of $a$. The Friedmann
equation (23) determines $a$ as a function of $G$ and $\Lambda$.
Then, the Bianchi identity (19) relates $G$ to $\Lambda$.

Let us consider the diffusion equation (28) in more detail in the
spatially flat FLRW model ($k=0$ in eq.(17)). In the metric (41)
the non-vanishing Christoffel symbols are
\begin{displaymath}
\Gamma^{0}_{jk}=a\frac{da}{dt}\delta_{jk},
\end{displaymath}\begin{displaymath}
\Gamma^{j}_{0k}=a^{-1}\frac{da}{dt}\delta^{j}_{k}\equiv
H\delta_{jk}.
\end{displaymath}

 Then, eq.(28) reads
\begin{equation}\begin{array}{l}
p_{0}(\partial_{t}-2Hp^{k}\partial^{p}_{k})\Omega+p^{k}\partial_{k}^{x}\Omega=\kappa^{2}
\Big(m^{2}a^{-2}\triangle_{p}+p^{j}p^{k}\partial^{p}_{j}\partial^{p}_{k}
+3p^{k}\partial^{p}_{k}\Big)\Omega,
\end{array}\end{equation}where $p_{0}=\sqrt{m^{2}+a^{2}{\bf p}^{2}}$ and ${\bf p}^{2}=p^{j}p^{j}$ .

 Let $r=\vert {\bf p}\vert$. Then, the formula   (49)
is
\begin{equation}\begin{array}{l}
\sqrt{m^{2}+a^{2}r^{2}}(\partial_{t}-2Hr\partial_{r})\Omega+p^{k}\partial_{k}^{x}\Omega=\kappa^{2}
\Big(m^{2}a^{-2}\triangle_{p}+r^{2}\partial_{r}^{2}
+3r\partial_{r} \Big)\Omega.
\end{array}\end{equation}On rotation invariant functions
$\triangle_{p}=\partial_{r}^{2}+\frac{2}{r}\partial_{r}$.
  The massless diffusion
is a soluble model in the spatially flat FLWR model ($k=0$ in
eq.(17)) which will be discussed in detail in sec.5. In this case
eq.(50) is
\begin{equation}\begin{array}{l}
a(\partial_{t}-2Hr\partial_{r})\Omega+p^{k}\vert{\bf
p}\vert^{-1}\partial_{k}^{x}\Omega=\kappa^{2} (r\partial_{r}^{2}
+3\partial_{r} )\Omega.
\end{array}\end{equation}

It can be seen from eq.(50) that if $H$ stays bounded  then the
limit $a\rightarrow \infty$ of the massive diffusion is the
massless diffusion.  As a consequence of the rotation invariance
of  $\triangle_{H}$ and $p^{j}\partial_{j}=r\partial_{r}$, the
diffusion operator in eq.(49) is rotationally invariant.
  If the initial state $\Omega$ is independent of ${\bf x}$ and rotationally invariant then the solution $\Omega_{t}$  depends
only on $\vert{\bf p}\vert$. We can calculate
\begin{equation}\begin{array}{l}
w=p\rho^{-1}=\frac{1}{3}a^{-2}\int d{\bf p}p_{0}^{-1}{\bf
p}^{2}\Omega_{t}\Big(\int d{\bf
p}p_{0}\Omega_{t}\Big)^{-1}\cr=\frac{1}{3}-\frac{1}{3}m^{2}\int
d{\bf q}\omega^{-1}\Omega_{t}(\int d{\bf q}\omega\Omega_{t})^{-1},
\end{array}\end{equation}where we set ${\bf q}=a{\bf p}$ and
$\omega=\sqrt{m^{2}+{\bf q}^{2}}$. Hence, $0<w\leq\frac{1}{3}$ is
time-dependent. It follows from eq.(52) that $w\rightarrow 0$ when
$m\rightarrow\infty $. In general,   $w$ could depend on ${\bf x}$
if $\rho$ and $P$ depend on ${\bf x}$. Eq.(42) with a time
dependent $w$ has the form
\begin{equation}
\partial_{t}\rho+3(1+w_{t})H\rho=3n\kappa^{2}a^{-3}.
\end{equation}
The solution of eq.(53) is discussed in the next section. Note
that inserting as $\Omega_{t}$ the stationary J\"uttner
distribution \cite{wert} of sec.5 with
 the temperature $\frac{\kappa^{2}}{H}$ \cite{haba0} we obtain
$w= \frac{1}{3}-\frac{1}{3}K$ with

\begin{equation}\begin{array}{l}
K=\int_{0}^{\infty}dy
y^{2}(m^{2}+y^{2})^{-\frac{1}{2}}\exp(-\frac{H}{\kappa^{2}}(m^{2}+y^{2})^{\frac{1}{2}})
\cr \Big(\int_{0}^{\infty}dy
y^{2}(m^{2}+y^{2})^{\frac{1}{2}}\exp(-\frac{H}{\kappa^{2}}(m^{2}+y^{2})^{\frac{1}{2}})\Big)^{-1}.
\end{array}\end{equation}

\section{Thermodynamics in an expanding universe}We consider a homogeneous thermodynamic system in an instantaneous equilibrium.
At the fixed time for a system in a finite volume $V$ we  consider
the first law of thermodynamics for the energy $E$ with the energy density $\rho$ (so $E=\rho V$) where  $\rho=\rho(T)$ depends only on temperature) and the entropy
$S=S(V,T)$ 
\begin{equation}
TdS=dE+PdV=\frac{\partial \rho}{\partial T}VdT
+(P+\rho)dV=T(\frac{\partial S}{\partial T}dT+\frac{\partial
S}{\partial V}dV),
\end{equation}(the differential is a difference between various
configurations).

Hence,
\begin{equation}
\frac{\partial \rho}{\partial T}V=T\frac{\partial S}{\partial T}.
\end{equation}

 We assume $P=w\rho$. 

Differentiating eq.(56)  over $T$ we get
\begin{equation}
\frac{\partial S}{\partial V}+T\frac{\partial^{2}S}{\partial
V\partial T}=(1+w)\frac{\partial \rho}{\partial T}.\end{equation}

Differentiating $dV$ component of eq.(55) over $V$ \begin{equation}
T\frac{\partial^{2}S}{\partial T\partial
V}=\frac{\partial\rho}{\partial T}.
\end{equation}
From eqs.(55),(57)-(58) we obtain
\begin{equation}
\frac{\partial\rho}{\partial T}=w^{-1}(1+w) \frac{\rho}{T}
\end{equation}
Hence, from eq.(59)
\begin{equation}
\rho_{st}(T) =BT^{\frac{1+w}{w}} \end{equation} and from eq.(55)
\begin{equation}
S_{st}=(1+w)BVT^{\frac{1}{w}},
\end{equation}where the index $st$ refers to a fixed time.  Let us consider how these equations will evolve in time.
 In the flat expanding metric $V=V_{0}a^{3}$( we
take $V_{0}=\frac{4\pi}{3}$ later on). Hence, if in a closed
(isolated) system  $S$ is to be constant, with a constant $B$ of
eq.(60), then $T$ must vary as $T=T_{0}a^{-3w}$ (a consequence of
eq.(61)).

We assume that in a dissipative adiabatic process in spite of the
dependence of $V=V_{0}a^{3}$ on time the entropy $S_{st}$ (61)
does not change. This gives the condition
\begin{equation}
(1+w)B(a)a^{3}T(a)^{\frac{1}{w}}=const
\end{equation}
If we know $\rho_{st}$ (as we got it in eq.(43)) then from eq.(60)
and (62) we determine $T(a)$ for the dissipative system. In the
case of the diffusion, eq.(43) together with eqs.(62) and (60)
define $T(a)$ in the form (44).

 Let us assume that  in the diffusive system in an
expanding universe the entropy is the sum
\begin{equation}
S=S_{st}+S_{diss},
\end{equation}where $S_{st}$ is the constant from eqs.(61)-(62).
 For the relativistic diffusion we obtain (from the equation
 $TdS=dE+PdV$, and eqs.(35)(42))
\begin{equation}\begin{array}{l}
T\partial_{t}S_{diss}=V(\partial_{t}\rho)+V^{-1}\partial_{t}V\rho(1+w)=V3\kappa^{2}na^{-3}
=3\kappa^{2}nV_{0},\end{array}
\end{equation}
where $w$ may depend on time.

Then, from eqs.(64) and (44)
\begin{equation}
S_{diss}=3\kappa^{2}nV_{0}\int_{t_{0}}^{t}\frac{ds}{T(s)}+C=3nV_{0}\ln(T_{0}+\kappa^{2}A)+C,
\end{equation}
where $C$ is a constant. Note that in eq (47) $\ln G\simeq -S$.

The formulas above apply to time-dependent $w$ with minor changes.
Let
\begin{equation}
a^{-3w_{e}(t,\tau)}\equiv
\exp(-3\int_{\tau}^{t}w_{s}\partial_{s}\ln a ds)
\end{equation}
Then, from eq.(42)
\begin{equation}\begin{array}{l}
\rho(t)=3na^{-3(1+w_{e}(t,t_{0}))}(T_{0}+\kappa^{2}\int_{t_{0}}^{t}a^{3w_{e}(\tau,t_{0})}d\tau)
\equiv 3na^{-3(1+w_{e}(t,t_{0}))}B(a).
\end{array}\end{equation} We define the temperature in such a way
that $S_{st}$ (62) does not change during the diffusive expansion.
Then, it can be seen from eq.(67) that the formula for the
temperature is
\begin{equation}\begin{array}{l}
T(t,a)=a^{-3w_{e}(t,t_{0})}B(a)=a^{-3w_{e}(t,t_{0})}(T_{0}+\kappa^{2}\int_{t_{0}}^{t}a^{3w_{e}(s,t_{0})}ds).
\end{array}\end{equation} In such a case the integration in eq.(65) gives
the formula for the dissipative entropy \begin{equation}
S_{diss}=3nV_{0}\ln\Big(T_{0}+\kappa^{2}\int_{t_{0}}^{t}a^{3w_{e}(s,t_{0})}ds\Big).
\end{equation}
In fact, $S_{diss}$ is the solution of the equation (64)
$\partial_{t}S_{diss}=3\kappa^{2}nT^{-1}V_{0}$.
\section{Solutions with the initial J\"uttner distribution}
We are unable to solve the massive equation (28) for a general
expansion $a$.
 In the de Sitter space  $a=\exp(Ht)$  the solution
with the J\"uttner initial distribution (obtained in
\cite{habafriction}) is
\begin{equation}\Omega_{t}=\frac{n}{8\pi} (\kappa^{2}H^{-1}a)^{-3}\exp(-
\sqrt{a^{2}{\bf p}^{2}+m^{2}}H\kappa^{-2}).
\end{equation}
It looks (up to a numerical time-dependent factor) as a static
J\"uttner distribution \cite{juttner} in the physical momentum
${\bf q}=a{\bf p}$. In a sense at any time we have again a J\"uttner-type distribution. It is expressing the fact that de Sitter space
has a definite temperature (here $T= \frac{\kappa^{2}}{H}$ \cite{haba0}). The
diffusion equilibrates to this temperature. In de Sitter space the
relativistic diffusion (28) can be expressed as a diffusion  with
a time-independent generator. In fact, let us introduce new
variables $ (a,u)$ with $u=a\vert{\bf p}\vert=ar$ then eq.(28)in
the $(t,u)$ variables for a rotationally
invariant initial condition reads
\begin{equation}\begin{array}{l}
\partial_{t}\Omega=\Big(Hu\partial_{u}+\frac{m^{2}\kappa^{2}}{\omega}
(\partial_{u}^{2}+\frac{2}{u}\partial_{u}) 
+\frac{\kappa^{2}}{\omega}(u^{2}\partial_{u}^{2}+3u\partial_{u})\Big)\Omega\equiv
{\cal L}\Omega,
\end{array}\end{equation}
From eq.(71) it follows that  J\"uttner distribution
\cite{juttner} $\exp(-H\kappa^{-2}\omega)$ where
$\omega=\sqrt{m^{2}+u^{2}}$ is the eigenfunction of ${\cal L}$
with the eigenvalue $-3H$ so that

$\exp(-3H\tau)\exp(-H\kappa^{-2}\omega)$ is the solution of
eq.(71). For a large $u$ the term proportional to the mass drops
out. Then, asymptotically we obtain from eq.(71) the solution
$\exp (-H\kappa^{-2}u$ ) of the massless diffusion (eq.(71) is
still discussed in sec.7).

 We
restrict ourselves in this section to the soluble model of $m=0$.
At $m=0$ eq.(28) reads ( where $r=\vert {\bf p}\vert$)
\begin{equation}\begin{array}{l}
a(\partial_{t}-2Hr\partial_{r})\Omega=\kappa^{2}
(r\partial_{r}^{2} +3\partial_{r} )\Omega.
\end{array}\end{equation}
This equation is exactly soluble (after a change of time) for arbitrary $a(t)$ and general initial conditions as will
be discussed in sec.7 and in the Appendix. However, in subsequent
two  sections we restrict our discussion to the solutions of this
equation with an initial J\"uttner distribution \cite{juttner} (it
has an interpretation as a relativistic Gibbs distribution \cite{mackey}).

Let
\begin{displaymath}
a{\bf p}={\bf q}.
\end{displaymath}
  We solve eq.(72) with the initial condition  (we set the Boltzman constant $k_{B}=1$, then defining
the temperature $\tilde{T_{0}}$ in Minkowski space we shall have
$k_{B}\tilde{T}_{0}=\frac{T_{0}}{a_{0}}$)
\begin{equation}
\Omega_{0}=\frac{n}{8\pi}T_{0}^{-3}\exp(-\frac{a_{0}\vert{\bf
q}\vert}{T_{0}}).
\end{equation}
It is normalized as
\begin{displaymath}
\int d{\bf q} \sqrt{g_{0}}\Omega_{0}=n\equiv\frac{N}{V_{0}},
\end{displaymath}where $N$ is the number of particles, $V_{0}$ is
the initial volume, so  that  $n$ is the initial particle density.
The solution of eq.(72) with the initial condition (73) at $t_{0}$
is \cite{habagrg}\cite{habampa}(we assume $a(t_{0})=a_{0})$
\begin{equation}\Omega_{t}=\frac{n}{8\pi} (T_{0}
+\kappa^{2}A)^{-3}\exp\Big(-\frac{a\vert{\bf
q}\vert}{T_{0}+\kappa^{2}A}\Big),\end{equation}where
\begin{equation}
A(t)=\int_{t_{0}}^{t}a(s)ds.
\end{equation}
The instantaneous distribution is again of J\"uttner-type.
It could be considered as a solution of the maximum entropy requirement \cite{mackey} for the entropy defined by the phase space distribution (see sec.6).
It is normalized
\begin{equation}
\int d{\bf q}\sqrt{g_{t}}\Omega_{t}=n.
\end{equation}
Note that the solution (74) in general has no limit (no
normalizable equilibrium) as $t\rightarrow \infty$. The exception
is the Sitter space ($a=\exp Ht$) when the solution (74)  becomes
time-independent (up to a normalization factor)and coincides with
(70) at $m=0$ if
\begin{displaymath}
T_{0}=H^{-1}\kappa^{2}a(t_{0}).
\end{displaymath}

The current $N^{\mu}$ is covariantly conserved as from eq.(28)

\begin{equation}
\nabla^{x}_{\mu}N^{\mu}=\frac{1}{\sqrt{g}}\partial^{x}_{\mu}(\sqrt{g}N^{\mu})=
a^{-3}\partial_{t}(a^{3}N^{0})=a^{-3}\partial_{t}n=0.\end{equation}
In fact, eq .(77) holds true in general because from the diffusion
equation (28)we can show $\nabla_{\mu}^{x}N^{\mu}=0$.

The normalization (76) is a fulfillment of the conservation law
\begin{equation}
\int_{V_{0}} d{\bf x}\sqrt{-g}N^{0}=\int_{V_{0}} d{\bf x}d{\bf
q}\sqrt{-g}\Omega=const
 \end{equation}Hence, $N^{0}=\frac{n}{\sqrt{-g}}$.

 For a fixed $T_{0}$ in the state (74) we calculate $\rho$ from
the phase space distribution in eq.(26)( $P=\frac{1}{3}\rho $ as a
consequence of eq.(26) and $p^{\mu}p_{\mu}=0$ )
\begin{equation}
\rho =T^{00}=3na^{-4}(T_{0}+\kappa^{2}A)
\end{equation}(in agreement with the solution (43) for
$w=\frac{1}{3}$).  Note that a  superposition
\begin{equation}
\Omega=\int dT_{0}\nu (T_{0})\Omega(T_{0})
\end{equation}
(with $\int dT_{0}\nu (T_{0})=1$) will also solve the diffusion
equation (28) with the same normalization as the initial J\"uttner
distribution. It has the energy density

\begin{equation}\begin{array}{l}
\rho =3na^{-4}\int
dT_{0}\nu(T_{0})(T_{0}+\kappa^{2}A)=3nCa^{-4}+3n\kappa^{2}a^{-4}A\cr=
3na^{-4}(\tilde{T}_{0}+\kappa^{2}A),\end{array}\end{equation} where
\begin{displaymath} \tilde{T}_{0}=\int dT_{0}\nu(T_{0})T_{0}.
\end{displaymath}

\section{Entropy variation}

In this section we suggest a covariant form of the first law of thermodynamics and statistical mechanics for a metric more
general than (17). For this purpose let us introduce the inverse
temperature $\beta$ as a four-vector (see \cite{israel}
\cite{wert}\cite{beca}\cite{beca2} and references cited there)
\begin{equation}
\beta^{\mu}=\beta u^{\mu}.
\end{equation}

 Then, we consider
the entropy density $\sigma^{\mu}$ as the four-vector
\begin{equation}
\sigma^{\mu}=\sigma u^{\mu}.
\end{equation}
We express the first law of  thermodynamics of a moving fluid as \cite{jhzh}
\begin{equation}
-\nabla_{\mu}\sigma^{\mu}=	\beta_{\nu}\nabla_{\mu}T^{\mu\nu}.
\end{equation}or\begin{displaymath}
\nabla_{\mu}(\beta_{\nu}T^{\mu\nu})-\frac{1}{2}(\nabla_{\mu}\beta_{\nu}+\nabla_{\nu}\beta_{\mu})T^{\mu\nu}=-\nabla_{\mu}\sigma^{\mu}.
\end{displaymath}
Hence, if $\beta_{\mu}$ is the Killing vector then

\begin{displaymath}
\sigma^{\mu}=-	\beta_{\nu}T^{\mu\nu}+k^{\mu},
\end{displaymath}where $\nabla_{\mu}k^{\mu}=0$. 
We can insert eq.(84) in the Bianchi identities (12)
\begin{displaymath}\begin{array}{l}
\partial_{t}G\int d{\bf
	x}\sqrt{\gamma}\beta T^{0}_{0}+G\partial_{t}\int d{\bf
	x}\sqrt{-g}\sigma^{0}+G \int d{\bf x}
\partial_{j}(\sqrt{-g}\sigma^{j})\cr+\frac{1}{8\pi}\int d{\bf
	x}\sqrt{\gamma}\beta \partial_{t}\Lambda=0.\end{array}
\end{displaymath}
In the comoving frame $\sigma^{j}=\sigma
u^{j}=0$. We define the entropy as
\begin{displaymath}
S=\int d{\bf x}\sqrt{\gamma}\sigma
\end{displaymath}and the mass
\begin{displaymath}
M\int d{\bf
	x}\sqrt{\gamma}T^{00}\end{displaymath}
Then, using  $\sigma^{0}=u^{0}\sigma=\vert
g_{00}\vert^{-\frac{1}{2}}\sigma$ we can rewrite this equation as
\begin{displaymath}
\begin{array}{l}\beta
\partial_{t}G M+G\partial_{t}S+\frac{\beta}{8\pi}\int
d{\bf x}\sqrt{\gamma} \partial_{t}\Lambda=0.
\end{array}\end{displaymath}
It is interesting to note that if we apply this equation to a system with the horizon $2GM$, the Hawking temperature $T$ and the Hawking formula for the entropy $S=\frac{4\pi (2GM)^{2}}{4G}$ then we get as a conclusion $G\simeq M^{-\frac{2}{3}}$.

For a diffusing fluid from eq.(35) we obtain
\begin{equation}
3\kappa^{2}\beta^{\nu}N_{\nu}=-\nabla_{\mu}\sigma^{\mu}.
\end{equation}
In the comoving frame (from eq.(85))
\begin{equation}
(-g)^{-\frac{1}{2}}\partial_{t}(\sigma^{0}\sqrt{-g})=3\kappa^{2}\beta_{0}N^{0}.
\end{equation}
With the temperature (44) eq.(86) reads
\begin{equation}
\partial_{t}(\sigma^{0}a^{3})=\partial_{t}\ln(T_{0}+\kappa^{2}\int^{t}_{t_{0}}
a^{3w}).
\end{equation}
Hence,
\begin{equation}
\sigma^{0}a^{3}=3n\ln(T_{0}+\kappa^{2}\int^{t}_{t_{0}} a^{3w})+C.
\end{equation}
(we assume that all variables depend only on time). We integrate
over an expanding volume $V=V_{0}\sqrt{g}$ ( we assume the
isotropic expansion  $\vert {\bf x}(t)\vert =a(t)$, $a(0)=1$,
$V_{0}=\frac{4\pi}{3}$ ) then
\begin{equation}
S=\int d{\bf x}\sqrt{g}\sigma=\sigma
\sqrt{g}V_{0}=\frac{4\pi}{3}\sigma a^{3}.
\end{equation}Hence, from eqs.(84)-(85) we recover our result
(69).

We still rewrite eq.(84)  using $\nabla_{\mu}u^{\mu}=3H$ (for FLRW metric (17))

 \begin{equation} \nabla_{\mu}\sigma^{\mu}
=u^{\mu}\nabla_{\mu}\sigma +\sigma\nabla_{\mu}u^{\mu}=
u^{\mu}\nabla_{\mu}\sigma +3H\sigma.
\end{equation}
In the comoving frame $u=(1,{\bf 0})$ we obtain from eq.(84)and
the energy-momentum tensor of an ideal fluid   the fluid
continuity equation
\begin{equation}
    \partial_{t}\rho+3H(\rho+\tilde{P})=T\partial_{t}\sigma,
\end{equation}
where $\tilde{P}=P-\sigma T$ and $\rho+\tilde{P}$ is the entalpy
density. Eq. (91) expressed in terms of the entropy (89) gives the
first law of thermodynamics (55)(with $V=\frac{4\pi}{3} a^{3})$
\begin{equation}
    \partial_{t}\rho+3H(\rho+P)=\frac{3}{4\pi a^{3}}T\partial_{t}S.
\end{equation}
Inserting the Bianchi identity (19) in eq.(92) we express it in
the form
\begin{equation}
8\pi\rho
\partial_{t}G +8\pi
TG(\nabla_{t}\sigma +3H\sigma)+\partial_{t}\Lambda=0.
\end{equation}
or in terms of the entropy $S$ (89)

\begin{equation}\begin{array}{l}
        a^{-3}GT\frac{dS}{dt}=-\frac{4\pi}{3}\frac{dG}{dt}\rho
        -\frac{1}{6}\frac{d\Lambda}{dt}.
\end{array}\end{equation}

For a diffusing system of particles having a phase space
distribution $\Omega$ we can calculate the statistical physics
(microscopic) entropy $\sigma^{\mu}$ from the formula
\cite{wert}\cite{lebowitz}
\begin{equation}
\sigma^{\mu}=-\sqrt{g}\int d{\bf p}p_{0}^{-1}p^{\mu}\Omega\ln
\Omega=\sigma u^{\mu}.
\end{equation}
In the comoving frame $\sigma=\sigma^{0}$. Then, from eqs.(74) and
(95)
\begin{equation}\begin{array}{l}
\sigma=-\sqrt{g}\int d{\bf p}\Omega\ln \Omega=-\int d{\bf
q}\Omega\Big(-(T_{0}+\kappa^{2}A)^{-1}\vert {\bf
q}\vert-3\ln(T_{0}+\kappa^{2}A)+\ln n\Big)\cr=
3na^{-3}\ln(T_{0}+\kappa^{2}A)-a^{-3}(n\ln n-3n).
\end{array}\end{equation}Then
\begin{equation}
S=\frac{4\pi}{3}sa^{3}=4\pi
n\ln(T_{0}+\kappa^{2}A)-\frac{4\pi}{3}(n\ln n-3n).
\end{equation}
Hence, in eq.(65)( with $V_{0}=\frac{4\pi}{3}$) we have
$C=\frac{4\pi}{3}(-n\ln n+3n)$. The statistical and thermodynamic
entropies coincide. When expressed by the temperature $S= 4\pi
n\ln (a^{3w}T)$, the result for the entropy resembles the
Sackura-Tetrode formula for a gas of free non-relativistic
particles.

For massive
particles in de Sitter space the formula (95) with the solution (70) gives\begin{equation}
S\simeq Kt
\end{equation}
It can be seen that such a result also follows from eq.(97) as
$\ln (T_{0}+A)\simeq t $.  For a power-law expansion we obtain a slower growth of
entropy $S\simeq \ln t$ at large time.

We have derived  the result  for the entropy of the gas of
diffusing particles first from thermodynamics knowing the formula
for the energy density. Then, this result has been obtained from a
microscopic Gibbs-type distribution as a solution of the diffusion
equation. Clearly, for other solutions of the diffusion equation
we would get from eq.(95) another formula for the entropy of this distribution
(eventually asymptotically as $t\rightarrow \infty$ we could approach
(97)).

\section{Solution of Einstein and diffusion equations}
We can get a general solution of the massless diffusion equation
(72) in a general spatially flat expanding metric because after a change of
variables it can be reduced to the equation in the Minkowski space
time solved in \cite{habampa}\cite{transport}.  Let in eq.(72) 
$v=a^{2}r=a^{2}\vert {\bf p}\vert$. We introduce a new time $\eta$

\begin{equation}
\kappa^{2}\eta=T_{0}+\kappa^{2}\int_{t_{0}}^{t}a(s)ds
\end{equation}
The variable $\ln \eta\simeq S$ related to the entropy $S$ in eq.(65) and to the temperature by eq.(44) could be considered as a thermal time \cite{connes} in general relativity
( in fact, in the state (74) the relation $\frac{d}{dt}=\frac{1}{T} \frac{d}{d\eta}$ of ref.\cite{rovelli} holds true with the temperature $T$   (44) although the state (74) is not stationary).
In the variables $(\eta, v)$  eq.(72) reads
\begin{equation}\begin{array}{l}
\partial_{\eta}\Omega+a^{-2}p^{k}\vert{\bf
p}\vert^{-1}\partial_{k}^{x}\Omega=\kappa^{2} (v\partial_{v}^{2}
+3\partial_{v} )\Omega.
\end{array}\end{equation}The momentum part of eq.(100) becomes time-independent.
For further purposes we consider a slightly more general equation
\begin{equation}\begin{array}{l}
\partial_{\eta}\Omega+a^{-2}p^{k}\vert{\bf
p}\vert^{-1}\partial_{k}^{x}\Omega=hv\partial_{v}\Omega+\kappa^{2}
(v\partial_{v}^{2} +3\partial_{v} )\Omega={\cal L}\Omega.
\end{array}\end{equation}
In the variable $y=-\kappa^{-2}h v$  Laguerre polynomials
\cite{gradstein} are solutions of the eigenvalue equation
\begin{equation}
{\cal L}L^{(2)}_{k}(y)=khL^{(2)}_{k}(y).
\end{equation}
It follows that $\exp(kh\eta)L^{(2)}_{k}(y)$ are solutions of the
diffusion equation (101). Hence, an arbitrary sum of these
functions is a solution. It is useful to apply the known formulas
for the summation of Laguerre polynomials
\cite{gradstein}\begin{equation} \sum_{k=0}^{\infty}
s^{k}L^{(\alpha)}_{k}(y)=(1-s)^{-\alpha -1} \exp(-\frac{sy}{1-s})
\end{equation}
and
\begin{equation}\begin{array}{l}
\sum_{k=0}^{\infty}n!\Gamma(k+\alpha+1)^{-1}L_{k}^{(\alpha)}(x)L_{k}^{(\alpha)}(y)s^{k}
\cr=(1-s)^{-1}(xys)^{-\frac{1}{2}\alpha}\exp\Big(-s(1-s)^{-1}(x+y)\Big)
I_{\alpha}(2(1-s)^{-1} \sqrt{xys}),
\end{array}\end{equation}
where $I_{\alpha}$ is the Bessel function. Using the orthogonality
property of Laguerre polynomials
\begin{equation}
n!\Gamma(k+\alpha+1)^{-1}\int_{0}^{\infty}dx
x^{\alpha}L_{n}^{(\alpha)}(x)L_{m}^{(\alpha)}(x)=\delta_{nm}
\end{equation}
we can obtain  a solution of the diffusion equation with any
initial condition.

Using eqs.(102)-(105) we can generate solutions of eq.(101). Then,
taking the limit $h\rightarrow 0$ we obtain a general solution of
eq.(100). From eq.(103) we obtain in the limit $h\rightarrow 0$
(up to a normalization) the solution (74)
\begin{equation}
\Omega_{\eta}(v)=\eta^{-3}\exp(-\kappa^{-2}\frac{v}{\eta})
\end{equation}
From eq.(104) we obtain an evolution kernel for eq.(101)\begin{equation}\begin{array}{l}
\Omega_{\eta}(v,v^{\prime})=K(vv^{\prime}\eta)^{-1}
\exp(-\frac{v+v^{\prime}}{\kappa^{2}\eta})I_{2}(\frac{2\sqrt{vv^{\prime}}}{\kappa^{2}\eta}),
\end{array}\end{equation}
where $K$ is a normalization constant such that when
$\eta\rightarrow 0$ then $\Omega_{\eta}(v,v^{\prime})\rightarrow
\delta(v-v^{\prime})$

The solution of the space-time dependent  diffusion equation (
with the initial condition $\Omega({\bf x},{\bf p}\vert{\bf
p}\vert^{-1},\vert {\bf p}\vert))$ reads

\begin{equation}\begin{array}{l}
\Omega_{\eta}({\bf x},{\bf p}\vert{\bf p}\vert^{-1},\vert {\bf
p}\vert)=E[\Omega({\bf x}-{\bf p}\vert{\bf p}\vert^{-1}\tau, {\bf
p}\vert{\bf
p}\vert^{-1},v_{\eta})]\cr=\int_{0}^{\infty}dv^{\prime}
\Omega({\bf x}-{\bf p}\vert{\bf p}\vert^{-1}\tau, {\bf p}\vert{\bf
p}\vert^{-1},v^{\prime}) \Omega_{\eta}(v,v^{\prime}),
\end{array}\end{equation} where $E[...]$ denotes the expectation value over the stochastic process
$v_{\eta}$ generated by the diffusion (101) (see the Appendix) and
$\tau=\int dta^{-1}$ is the conformal time,
$\Omega_{\eta}(v,v^{\prime})$ is the evolution kernel (107) (the
diffusion transition function).

The phase space solution (108) could be used for a calculation  of space and momentum evolution of a cluster of particles.
However, it is not simple to apply eqs.(26)-(27) for analytic calculations of $\hat{T}^{\mu\nu}$ and the current $N^{\mu}$ of eqs. (26)-(27). 
The solution (108)  is  heavily dependent on the initial
condition and it is not so explicit concerning its dependence on
the cosmic time $t$. For this reason let us discuss some simple
solutions based on the equations for $\rho$ and $a$.
 If $G=const $ then the Bianchi identity (19) (or
(45)), the expression of the energy-momentum non-conservation (42)
and the first Friedmann equation (23) constitute a closed system
of first order differential equations for $\Lambda,\rho, a$ with
the unique solution under given initial conditions ( in
particular, we may assume for $\Lambda$ zero or non-zero initial
condition). If we admit $G$ to depend on time then in order to
have the unique solution we must choose  one of the functions
$a,\rho,\Lambda$ or $G$ .

We can look for a solution of the diffusion equation (28) as a
part of a more ambitious cosmological program of a description of the
formation and time evolution of galaxies in the universe
\cite{galaxies} by means of their phase-space distribution
$\Omega_{t}({\bf x},{\bf p})$. Such a project requires a solution
of Einstein equations (2) together with the solution of  the
diffusion equation (28) (for other applications of the diffusion's phase space distribution see \cite{london1}\cite{london2}\cite{cruz}). We may begin with a simple task of
solving the diffusion equation with a fixed metric $g_{\mu\nu}$
which eventually can also solve the Einstein equations (2). We
define $\rho$ and $P$ by eq.(26) (as a function of $a$). Then, the
Friedmann equation (23) and the Bianchi identities determine two
of the three functions $a,G,\Lambda$.

As a simple example we consider $G=const$ and look for a
(asymptotic) power-law solution
\begin{equation}
a(t)=a_{1}t^{\alpha}.
\end{equation}From eq.(43)
\begin{equation}
\rho_{t}=Ct^{-3\alpha(1+w)}+At^{-3\alpha+1}.
\end{equation}
Then, from the Friedmann equation (23) we obtain
  $\alpha=1$ (coasting solution
\cite{kolb}) and ($C=0$ in eq.(110) or $w=-\frac{1}{3}$)
\begin{equation}
\rho(t)=At^{-2}
\end{equation}
\begin{equation} \Lambda=Bt^{-2}
\end{equation}
Eqs.(109)-(112) (with $\alpha=1$ in eq.(109) and $C=0$ in
eq.(110)) deliver a particular exact solution of the eqs.(19)(23)
and (42) of the diffusive gas (eq.(42) is a consequence of the
diffusion equation (28)). It can be seen from the derivation that
under the assumption $G=const$ the solution (109) with $\alpha=1$
is the asymptotic power-law solution of eqs.(19)(23) and (42)
(there may be an asymptotic exponential solution of these
equations if the initial condition for $\Lambda$ is a positive
number).

With $\rho$ and $\Lambda$ of eqs.(109)-(112) we still must determine the phase
space distribution $\Omega$ as the solution of the diffusion
equation (28).
 There may be various
solutions of the diffusion equation (28) (with different initial conditions) leading to the
energy-momentum and the current satisfying eqs.(19),(23) and (42)
(with the solution (109)-(112)). A particular solution of the
diffusion equation which leads to Einstein equations with
$a(t)=a_{1}t$, $\rho$ (111) and $\Lambda$ (112) is the solution of
eq.(72) for massless particles
\begin{equation}
\Omega_{t}=t^{-6}\exp(-2\kappa^{-2}\vert{\bf p}\vert).
\end{equation}
This solution follows from eq.(74) with a special choice of
parameters $T_{0}$, $t_{0},a_{1}$ in such a way that
\begin{displaymath}
\kappa^{2}\eta=T_{0}+\kappa^{2}\int_{t_{0}}^{t}a_{1}sds=\frac{\kappa^{2}}{2}t^{2}.
\end{displaymath}
We could find a solution of eqs.(19)(23) and (45) with an
arbitrary $\alpha$ in eq.(109), if $G$ is time-dependent. If the
solution for evolution equations is searched  in the form of a
power-law then there are power-law ( in cosmic time) components
for $a$ $\rho$, $G$ and $\Lambda$.

  The
coasting evolution $a(t)=a_{1}t$ is exceptionally simple.  In
particular, with a rotation invariant space independent initial
condition , if we introduce the conformal time $\tau=\int dt
a^{-1}=\ln t$, then eq.(72) can be expressed as (with $r=\vert
{\bf p}\vert$)
\begin{equation}
\partial_{\tau}\Omega=2r\partial_{r}\Omega+\kappa^{2}(r\partial_{r}^{2}
+3\partial_{r})\Omega={\cal L}\Omega.\end{equation} This is a
special case of eq.(101) ($h=2$).
The operator ${\cal L} $ in the variable $y=-2\kappa^{-2}r$ is
related to the operator defining the Laguerre polynomials
$L^{(\alpha)}_{k}$ with $\alpha=2$ . In fact,
\begin{equation}
{\cal L}L^{(2)}_{k}(y)=2kL^{(2)}_{k}(y).
\end{equation}
It follows that the functions $\exp(2k\tau)L^{(2)}_{k}$ are
solutions of eq.(114). Hence, also the sum of these functions is a
solution. Using the summation formula (103) for the Laguerre
polynomials we obtain a normalizable solution of the diffusion
equation (114) ($\tau$ is the conformal time)\begin{equation}
\Omega_{\tau}=(\exp(2\tau)-1)^{-3}\exp\Big(-\frac{2\exp(2\tau)}{\exp(2\tau)-1}\kappa^{-2}r\Big).
\end{equation}
The limit $\tau\rightarrow \infty$ of eq.(116) is also a solution
of eq.(114). It coincides with (113).

Using eq.(26) we can calculate the energy density for the solution
(116) . It is (in agreement with eq.(79))
\begin{equation}
\rho_{t}=At^{-2}+Ct^{-4}
\end{equation}
with certain constants $A$ and $C$.

We can obtain a general solution of eq.(114) with an arbitrary
initial condition $\Omega$. For this purpose we expand the initial
$\Omega$ in Laguerre polynomials $\Omega=\sum c_{k}L_{k}^{(2)}$.
Then, the solution of eq.(114) expressed as an infinite sum is
\begin{equation}
\Omega_{\tau}(y)=\sum_{k=0}^{\infty}c_{k}\exp(2k\tau)L_{k}^{(2)}(y).
\end{equation}
 The solution of
the space-time diffusion with the initial condition $\Omega({\bf
x},{\bf p}\vert{\bf p}\vert^{-1},\vert{\bf p}\vert)$ is (see the
Appendix)\begin{equation}\begin{array}{l} \Omega_{\tau}({\bf
x},{\bf p}\vert{\bf p}\vert^{-1},\vert{\bf
p}\vert)=\int_{0}^{\infty}dr^{\prime}\Omega({\bf x}-{\bf
p}\vert{\bf p}\vert^{-1}\tau,{\bf p}\vert{\bf
p}\vert^{-1},r)\Omega_{\eta}(r,r^{\prime}),\end{array}
\end{equation}
where $r=\vert {\bf p}\vert$, $\tau$ is the conformal time, $\Omega_{\eta}(r,r^{\prime})$ is the
evolution kernel of the diffusion operator (114) (the transition
function for the diffusion (114) defined in eq.(107)). The
elementary dependence of the phase space distribution on ${\bf x}$
is a consequence of eq.(49) as it follows from this equation that
${\bf p}\vert{\bf p}\vert^{-1}$ does not change during the
diffusion (see  the Appendix and
\cite{habampa}\cite{transport}).

As a next simple example of a solution   of a diffusion equation
(28) in a cosmic time let us consider an exponential expansion
$a=\exp(tH)$ ($H=const$). Eq.(28) can be expressed as a diffusion
with a time-independent generator. Let $u=ra$ then eq.(50) takes
the form

\begin{equation}\begin{array}{l}
\partial_{t}\Omega=Hu\partial_{u}\Omega+\kappa^{2}(m^{2}+u^{2})^{-\frac{1}{2}}
\Big(m^{2}(\partial_{u}^{2}+\frac{2}{u}\partial_{u})+u^{2}\partial_{u}^{2}
+3u\partial_{u} \Big)\Omega.
\end{array}\end{equation}
For large $u$ the $m$-dependent terms are of the order $mu^{-1}$
in eq.(120). Hence, the solution of eq.(120) for a large $u$
should not depend on $m$. Neglecting the $mu^{-1}$ terms we obtain
the equation
\begin{equation}\begin{array}{l}
\partial_{t}\Omega=Hu\partial_{u}\Omega +\kappa^{2}(u\partial_{u}^{2}
+3\partial_{u})\Omega\equiv {\cal L}\Omega.
\end{array}\end{equation}
We have obtained eq.(101) but now in the cosmic time. Introducing the
variable $y=-\kappa^{-2}H u$ we can see that Laguerre polynomials
are  solutions of the eigenvalue equation
\begin{equation}
{\cal L}L^{(2)}_{k}(y)=kHL^{(2)}_{k}(y).
\end{equation}
We may take the sum of Laguerre polynomials using the generating
function (103) in order to obtain a normalizable solution of the
diffusion equation (121)
\begin{equation}\begin{array}{l}
\Omega_{t}({\bf
p})=(\exp(Ht)-1)^{-3}\cr\exp\Big(-H\kappa^{-2}\vert{\bf p}\vert
\exp(2Ht)(\exp(Ht)-1)^{-1}\Big)\end{array}
\end{equation}(we skip the normalization constant).
The limit $t\rightarrow \infty$ of (123) is also the solution of
eq.(121). It coincides with the exact solution (70) of eq.(121) in
the $mu^{-1}$ expansion.
As in the general expansion $a(t)$ (108) and  in the case of the
coasting expansion we can obtain the general solution of the
massless equation expanding the initial condition (the momentum
part) $\Omega({\bf x},{\bf p}\vert{\bf p}\vert^{-1},\vert{\bf p}\vert)$ in Laguerre
polynomials by means of the orthogonality formula (105)( another
method is to apply the propagation kernel
$\Omega_{\eta}(v,v^{\prime})$ of eq.(107)). We can also express
the evolution kernel $\exp(\eta {\cal L})(r,r^{\prime})$ by
Laguerre polynomials using eq.(104). Then, the general solution of
eq.(121) can be expressed in the form similar to  (108). A
probabilistic expression of the solution (108) as well as (119) is
discussed in the Appendix.

As a next example of a particular solution in the cosmic time of
the diffusion equation let us consider eq.(28) in a flat expanding
metric with a large mass $m$ ( we neglect the terms of order
$\frac{1}{m^{2}}$ and $\frac{\kappa^{2}}{m}$ in eq.(49)). Then,
this equation reads
\begin{equation}\begin{array}{l}(\partial_{t}-2H(t)p^{k}\partial^{p}_{k})\Omega+m^{-1}p^{k}\partial_{k}^{x}\Omega
=\kappa^{2} ma^{-2}\triangle_{p}\Omega.
\end{array}\end{equation}
Eq.(124) describes a diffusion in momentum space with a
time-dependent diffusion constant. Its solution can be expressed
by the Brownian motion. This is the Gaussian stochastic process
defined by the variance ($E[..]$ denotes the expectation value
with respect of the Brownian motion)
\begin{equation}
E[b^{j}(t)b^{k}(s)]=min(t,s)\delta^{jk}.
\end{equation}
Then, the solution of eq.(124) with an initial condition
$\Omega({\bf x},{\bf p})$ at $t_{0}$ can be expressed as
\begin{equation}
\Omega_{t}({\bf x},{\bf p}))=E[\Omega({\bf x}-(t-t_{0})m^{-1}{\bf p},{\bf
p}(t))],
\end{equation}
where
\begin{equation}\begin{array}{l}
{\bf p}(t)=\exp(\int_{t_{0}}^{t}ds2H(s)){\bf p} + \kappa
\sqrt{2m}
\int_{t_{0}}^{t}\exp(\int_{s}^{t}ds^{\prime}2H(s^{\prime}))a(s)^{-1}d{\bf
b}(s).\end{array}
\end{equation}
We can obtain explicit integral formulas for the phase space evolution of the solution (127) if
we represent the initial condition $\Omega({\bf x},{\bf p})$ by
means of its Fourier transform.  Then, the expectation value over
the Brownian motion in eq.(126) can be calculated, e.g., the expectation value of its
Fourier factor is
\begin{equation}\begin{array}{l}
E[\exp\Big(\kappa \sqrt{2m} {\bf
y}\int_{t_{0}}^{t}\exp\int_{s}^{t}ds^{\prime}2H(s^{\prime}))a(s)^{-1}d{\bf
b}(s)\Big)]\cr= \exp\Big(\exp\Big(m\kappa^{2}{\bf
y}^{2}\int_{t_{0}}^{t}ds\exp\int_{s}^{t}ds^{\prime}4H(s^{\prime}))a(s)^{-2}\Big)\Big).
\end{array}\end{equation}

 With our result on the solution (123)
of the diffusion equation (121) for an exponential expansion we can approach the problem of the
solution of Einstein  equations assuming that the presumed exponential
expansion is an exact solution (as discussed earlier in order to
obtain the unique solution of Einstein  equations for $(a,\rho,G,
\Lambda)$ one of these functions must be chosen deliberately, here
we discuss the choice of $a$ as in (121))
\begin{equation} a(t)=a_{0}\exp(Ht)
\end{equation}

For the scale factor $a(t)$ (129) we can calculate $\rho_{t}$
using eq.(43) with the approximate constant value of $w_{t}$ from
eq.(52) (and its approximation at the end of sec.3, $w\simeq 0$ for $m\rightarrow \infty$)
\begin{equation}
\rho_{t}=A\exp(-3H(1+w)t)+B\exp(-3Ht).
\end{equation}

 The
Friedmann equation (24) gives
\begin{equation} 3H^{2}=\Lambda-4\pi G(1+3w)\rho.
\end{equation}
Inserting $\Lambda$ (131) in Bianchi identity (19) we obtain
\begin{equation}
\frac{3}{2}(1+w)\rho\partial_{t}G
+\frac{1}{2}(1+3w)\partial_{t}\rho G+3n\kappa^{2}Ga^{-3}=0
\end{equation}We set $\rho$ of eq.(130) in eq.(132) to obtain a linear equation for
$G$ with the solution
\begin{equation}\begin{array}{l}
G(t)=G(t_{0})(A\exp(-3H(1+w)t)+B\exp(-3Ht))^{-\nu}
\cr(A\exp(-3H(1+w)t_{0})+B\exp(-3Ht_{0}))^{\nu}\cr
(A+B\exp(3wHt))^{-\beta}(A+B\exp(3Hwt_{0}))^{\beta},
\end{array}\end{equation}
where
\begin{equation}
\nu=\frac{1+3w}{3(1+w)}
\end{equation}
and
\begin{equation}
\beta=2\kappa^{2}na_{0}^{-3}(3Hw(1+w))^{-1}.
\end{equation}
For a large $t$ and $w\geq 0$ it follows from eq.(133) that
\begin{equation}
G\simeq\exp(3Ht(\nu-w\beta))
\end{equation}
and
\begin{equation}
G\rho\simeq \exp(-3H((1-\nu)+w\beta)t)
\end{equation}
$\Lambda$ can be calculated from eq.(131). $\Lambda$ is decreasing
in time to its de Sitter value $3H^{2}$. It is noteworthy that an
exact exponential expansion leads to a growing Newton constant
(136). The diffusion responsible for a varying $G$ modifies $\rho$
of eq.(130) to $G\rho$ (137) entering the Friedman equation (23).

 Summarizing, in comparison to standard Friedmann 
equations (without diffusion) we have an additional equation relating $\nabla T$ with
a conserved current $N$.  This equation can be applied to express
$\rho$ by $a$. Then, the Bianchi identity is relating $G$ to
$\Lambda$.

The equations (44), (42) and (24) can be solved for particles with
$m>0$. Then, the index $w$ is time dependent. In such a case we do
not expect exact power-law solutions. However, the solutions with
constant $w$ can supply good approximations ( e.g. small $w\simeq
0$ for heavy particles and , $w\simeq \frac{1}{3}$ for light
particles).

\section{Summary}
We studied the thermodynamics and statistical mechanics of a gas diffusing in Einstein expanding universe.
The model can find applications to a gas of galaxies expanding in a medium of an unknown origin.
The thermodynamics is described by the energy density, pressure, entropy and temperature. 
In the diffusive model the covariant derivative of the energy-momentum is proportional to a conserved particle current.
The current allows a definition of the energy-momentum of an ideal fluid. For such systems we derive equations relating the energy-momentum to entropy and temperature.
The thermodynamics of the model is determined by the covariant derivative of the energy-momentum tensor. The statistical mechanics
results from the phase space distribution of diffusing particles. The phase space distribution can be obtained as a solution of the diffusion equation.
The statistical mechanics of a non-relativistic diffusive motion on a static Riemannian
manifold is well-known. We extend the statistical description to a relativistic diffusion on an
expanding universe. We have obtained the general solution of the relativistic diffusion equation for  massless particles. 
The massless equation is the limit of the massive diffusion equation for a large time.
Most interesting for a statistical description are the solutions with J\"uttner initial distribution
which can be obtained from the maximum entropy requirement.
The solutions determine the energy-momentum and the particle current of an ideal fluid.
The Einstein equations require that the covariant derivative of the total energy-momentum on the rhs
of Einstein equations is zero. We express this energy-momentum by  the
cosnmological term and the energy-momentum of an ideal fluid multiplied by the Newton "constant". In this way a variation of the cosmological constants can be expressed by a change of entropy.
We show that  the thermodynamic entropy  coincides with the microscopic entropy defined by the phase space distribution resulting as the solution of the diffusion equation
with the J\"uttner initial distribution.  We show that the entropy
can be considered as a distinguished time parameter which makes the coefficients in the diffusion equation in an expanding universe time-independent.
\section{Appendix}We begin with a simple stochastic description of the diffusion for the coasting
expansion which is an asymptotic solution of diffusive Einstein
equations. The solution of the diffusion equation (49) for $m=0$
and $a=t$ can be expressed by the solution of the stochastic
equations in the conformal time ($\tau=\ln t$ ) in the
Stratonovitch form \cite{ikeda}
\begin{equation}
d{\bf x}=-{\bf p}\vert{\bf p}\vert^{-1}d\tau,
\end{equation}

\begin{equation}
d{\bf p}(\tau)=3\kappa^{2}{\bf p}\vert{\bf p}\vert^{-1}d\tau+2{\bf
p}d\tau -\frac{1}{2}\kappa^{2}{\bf p}\vert{\bf
p}\vert^{-1}d\tau+\sqrt{2}\kappa {\bf p}\vert{\bf
p}\vert^{-\frac{1}{2}}\circ db(\tau).
\end{equation}
The Ito and Stratonovitch (denoted by the circle) differentials
\cite{ikeda} are related as

\begin{displaymath}
\vert{\bf p}\vert^{-\frac{1}{2}}p^{k}\circ db =\vert{\bf
p}\vert^{-\frac{1}{2}}p^{k}db+\partial_{j}(\vert{\bf
p}\vert^{-\frac{1}{2}}p^{k})dp^{j}db \end{displaymath}

From eq.(139) we can obtain (by differentiation) that $d({\bf
p}\vert{\bf p}\vert^{-1})=0$. Hence, ${\bf p}\vert{\bf
p}\vert^{-1}$ is equal to its initial value. We can also calculate
using eq.(139) \begin{equation} d\vert{\bf p}\vert= 3\kappa^{2}
+2\kappa^{2}\vert{\bf p}\vert d\tau
-\frac{1}{2}\kappa^{2}d\tau+\kappa\sqrt{2\vert{\bf p}\vert}\circ
db.
\end{equation}
Hence, it follows  that with $r=\vert{\bf p}\vert$ the diffusion
equation (51) for the coasting expansion $a=t$  is satisfied by
\begin{equation}
\Omega_{\tau}(\vert{\bf p}\vert)=E[\Omega(\vert{\bf p}_{\tau}\vert)].
\end{equation}
From the stochastic equations (138)-(141) and the consequence of
eq.(139) that ${\bf p}\vert{\bf p}\vert^{-1}$ is time independent
it follows that the solution of eq.(49) for $m=0$ with a spatial
dependence of $\Omega$ can be expressed as
\begin{equation}
\Omega_{\tau}({\bf x},{\bf p}\vert{\bf p}\vert^{-1},\vert {\bf
p}\vert))=E[\Omega({\bf x}-{\bf p}\vert{\bf p}\vert^{-1}\tau, {\bf
p}\vert{\bf p}\vert^{-1},\vert{\bf p}_{\tau}\vert)].
\end{equation}
The stochastic process $r_{\tau}=\vert{\bf p}_{\tau}\vert $ is the
solution either of the Stratonovitch equation (140) or the Ito
equation \begin{equation} dr=3\kappa^{2}d\tau+2rd\tau+\sqrt{2r}db
\end{equation}
There are minor changes in the derivation of the solution of
eq.(121) for the exponential expansion. The formula (142) still
applies in the form
\begin{equation}
\Omega_{t}({\bf x},{\bf p}\vert{\bf p}\vert^{-1},\vert {\bf
p}\vert))=E[\Omega({\bf x}-{\bf p}\vert{\bf p}\vert^{-1}\tau, {\bf
p}\vert{\bf p}\vert^{-1},u_{t})],
\end{equation}
where $u_{t}$ is the solution of the Ito equation
\begin{equation}
du=Hudt+3\kappa^{2}dt+\kappa\sqrt{2u}db(t).
\end{equation}
For an alternative study of the relativistic diffusion (28) it is
useful to derive its connection with a soluble model of quantum
mechanics of an anharmonic oscillator with the Hamiltonian
\cite{habareports}
\begin{equation}
H=-\frac{1}{2}\partial_{x}^{2}+\frac{1}{2}mx^{2}+gx^{-2}.
\end{equation}
The eigenvalues of this Hamiltonian are
$\epsilon_{n}=2mn+m(\alpha+1)$ where
\begin{displaymath}
\alpha=\frac{1}{2}\sqrt{1+8g}.
\end{displaymath} The eigenfunctions can be expressed by Laguerre
polynomials of order $n$. Let $\psi_{0}$ be the lowest eigenstate
of $H$ then
\begin{equation}
{\cal
L}=\psi_{0}^{-1}H\psi_{0}=-r\partial_{r}^{2}-(\alpha+1)\partial_{r}
+2mr\partial_{r}
\end{equation}
where $r=\frac{1}{2}x^{2}$. In eq.(147) (with $\alpha=2$ ) we
obtain the diffusion generator of eq.(101).

Returning to the general massless equation (51) in the time variable $\eta$
(101) and $v=a^{2}r$ we have

\begin{equation}\begin{array}{l}
\partial_{\eta}\Omega+a^{-2}p^{k}\vert{\bf
p}\vert^{-1}\partial_{k}^{x}\Omega=\kappa^{2} (v\partial_{v}^{2}
+3\partial_{v} )\Omega.
\end{array}\end{equation}
The original stochastic process for ${\bf x}$ and ${\bf p}$ reads
\begin{equation}
d{\bf x}=-{\bf p}\vert{\bf p}\vert^{-1}d\tau
\end{equation}
where $\tau$ is the conformal time $\tau=\int a^{-1}$ and

\begin{equation}
d{\bf p}(\eta)=3\kappa^{2}{\bf p}\vert{\bf p}\vert^{-1}d\eta+h{\bf
p}d\eta -\frac{1}{2}\kappa^{2}{\bf p}\vert{\bf
p}\vert^{-1}d\eta+\sqrt{2}\kappa {\bf p}\vert{\bf
p}\vert^{-\frac{1}{2}}\circ db(\eta),\end{equation}
where $\eta$ is defined in eq.(99).
It follows that  the solution of the diffusion equation (51) for
an arbitrary evolution $a(t)$ can be expressed by eq.(108).

\end{document}